\documentclass[twocolumn,showpacs,preprintnumbers,amsmath,amssymb,superscriptaddress,floatfix,nofootinbib]{revtex4}

\usepackage{graphicx}
\usepackage{amsmath}
\usepackage{amsfonts}
\usepackage{amssymb}

\begin{document}

\title{The $\bar{p} p \to \phi \phi $ reaction in an effective Lagrangian approach }
\author{Ju-Jun Xie} \email{xiejujun@impcas.ac.cn}
\affiliation{Institute of Modern Physics, Chinese Academy of
Sciences, Lanzhou 730000, China} \affiliation{Research Center for
Hadron and CSR Physics, Institute of Modern Physics of CAS and
Lanzhou University, Lanzhou 730000, China} \affiliation{State Key
Laboratory of Theoretical Physics, Institute of Theoretical Physics,
Chinese Academy of Sciences, Beijing 100190, China}
\author{Li-Sheng Geng}
\affiliation{School of Physics and Nuclear Energy Engineering and
International Research Center for Nuclei and Particles in the
Cosmos, Beihang University, Beijing 100191, China}
\author{Xu-Rong Chen}
\affiliation{Institute of Modern Physics, Chinese Academy of
Sciences, Lanzhou 730000, China}

\begin{abstract}

We investigate the $\bar{p} p \to \phi \phi$ reaction within an
effective Lagrangian approach.We show that the inclusion of either a
scalar meson $f_0$ or a tensor meson $f_2$ in the $s$-channel can
lead to a fairly good description of the bump structure of the total
cross section around the invariant $\bar{p} p$ mass $W \simeq 2.2$
GeV,  which cannot be reproduced with only the ``background"
contributions from $t$- and $u$-channel $N^*(1535)$ resonance as
studied in a previous work. From the fits, we infer the properties
of the involved scalar or tensor resonances.

\end{abstract}

 \maketitle

{\it Introduction}. According to the naive constituent quark model,
the $\phi$ meson is believed to be an almost pure $s\bar{s}$
state,~\footnote{In the quark model, the physical isoscalars $\phi$
and $\omega$ are mixtures of the SU(3) wave function $\psi_8$ and
$\psi_1$:
\begin{eqnarray}
\phi    &=& \psi_8 {\rm cos}\theta - \psi_1 {\rm sin}\theta \, , \\
\omega  &=& \psi_8 {\rm sin}\theta + \psi_1 {\rm cos}\theta \, ,
\end{eqnarray}
where $\theta$ is the nonet mixing angle and:
\begin{eqnarray}
\psi_8 &=& \frac{1}{\sqrt{6}} (u\bar{u} + d \bar{d} -2 s \bar{s}) \,
, \\
\psi_1 &=& \frac{1}{\sqrt{3}} (u\bar{u} + d \bar{d} + s \bar{s}) \,
.
\end{eqnarray}
For ideal mixing, ${\rm tan}\theta = 1/\sqrt{2}$ (or $\theta =
35.3^0$), the $\phi$ meson becomes pure $s\bar{s}$ state. } while
there are only up and down quarks (antiquarks) in the nucleon
(antinucleon). Thus the $\bar{p} p \to \phi \phi$ reaction, with its
disconnected quark lines, should be suppressed according to the
Okubo-Zeig-Iizuka (OZI) rule~\cite{ozirule}. However, even the OZI
rule is strictly enforced by nature, the $\bar{p}p$ reaction can
still proceed through the non-strange quark component of the $\phi$
meson, because of the slight discrepancy from the ideal mixing of
the vector meson singlet and
octet~\cite{pdg2012}.~\footnote{Experimentally, the mixing angle
$\theta$ is $36.4^0$. } With this small discrepancy, one can
determine an upper limit for the total cross section of $\bar{p} p
\to \phi \phi$ reaction by comparison to the total cross section of
the related $\bar{p} p \to \omega \omega$ reaction. This yields a
cross section for $\bar{p}p \to \phi \phi$ at the order of $10$
nb~\cite{Ellis:1994ww}. However, the experimental result from the
JETSET collaboration~\cite{Bertolotto:1994nz,Evangelista:1998zg}
showed that the cross section at $1.2$ GeV incident anti-proton
momentum, $\sigma= 2.86 \pm 0.46\,\mu b$,  is two orders of
magnitude larger than the estimated. Hence, the $\bar{p}p \to \phi
\phi$ reaction has attracted much attention because of the large OZI
rule violation~\cite{Ellis:1988jwa,Zou:1996ea}.

The large OZI violation has been interpreted by considering the
glueball candidate states which could break down the OZI
suppression~\cite{Lindenbaum:1980rf,Etkin:1982bw,Etkin:1987rj}, four
quark states containing a sizable $\bar{s}s$
admixture~\cite{Dover:1989zs} and the instanton induced interaction
between quarks~\cite{Kochelev:1995kc}. In
Refs.~\cite{Ellis:1988jwa,Ellis:1994ww,Ellis:1999er}, considerable
admixture of $s\bar{s}$ components in the nucleon was proposed to
explain the large OZI violation in $\bar{p}p$ annihilation. As a
result, it is often advocated that study of the $\bar{p}p \to \phi
\phi$ reaction could yield valuable information on the strangeness
content of the nucleon and nucleon resonances. On the other hand,
the large cross section for the $\bar{p}p \to \phi \phi$ reaction
could be explained by considering the two-step hadronic loops in
which each individual transition is
OZI-allowed~\cite{Lipkin:1984sw,Lipkin:1986bi}. Based on this, the
role played by two-meson ($\bar{K}K$) and antihyperon-hyperon
($\bar{\Lambda}\Lambda$) intermediate states in the $\bar{p} p \to
\phi \phi$ reaction have been studied by Lu ${\it
et~al.}$~\cite{Lu:1992xd} and Mull ${\it et~al.}$~\cite{mull},
respectively. All the aforementioned models are able to predict the
order of magnitude of the cross section, but not the detailed shape
of the observed spectrum~\cite{Evangelista:1998zg}, where there is a
bump around the invariant $\bar{p}p$ mass $W \simeq 2.2$ GeV, which
might hint at a sizable contribution from a scalar or tensor meson
in the $s$-channel.

Recently, Shi et al.~\cite{Shi:2010un} extended the work of
Ref.~\cite{Xie:2007qt}  to study the $\bar{p} p \to \phi \phi$
reaction by including the contributions from the $N^*(1535)$
resonance in the $t$- and $u$-channel. They showed that this new
mechanism may give significant contributions to the $\bar{p}p \to
\phi \phi$ reaction, especially for the invariant $\bar{p}p$ mass
$W$ above $2.3$ GeV. However, the bump structure below $2.3$ GeV
could not be reproduced, hinting at the necessity of including
contributions from the $s$-channel.

In the present work, we reanalyze the $\bar{p} p \to \phi \phi$
reaction within an effective Lagrangian approach and the isobar
model. In addition to the ``background" contributions from the
$N^*(1535)$ resonance studied in Ref.~\cite{Shi:2010un}, we propose
to introduce  s-channel contributions via either a scalar meson
$f_0$ or a tensor meson $f_2$.  Given the fact that the information
about the $f_0$ and $f_2$ meson with mass around $2.2$ GeV is
scarce~\cite{pdg2012}, we take the masses, the total decay width,
and the coupling constants, $g_{f_0\bar{p}p}g_{f_0 \phi \phi}$ and
$g_{f_2\bar{p}p}g_{f_2 \phi \phi}$ as free parameters, which will be
fitted to the experimental data on the $\bar{p} p \to \phi \phi$
reaction~\cite{Evangelista:1998zg}. In this respect, we show in this
work how the experimental study on the $\bar{p} p \to \phi \phi$
reaction may lead to the discovery of a strangeness scalar or tensor
resonance, or both around $2.2$ GeV.

This paper is organized as follows. In Section II, we present the
formalism and ingredients of our calculation. Numerical results and
discussions are given in Section III, followed by a short summary in
Section IV.

\section{Formalism and ingredients}

The effective Lagrangian method is an important theoretical tool in
describing the various processes around the resonance region. In
this section, we introduce the theoretical formalism and ingredients
to study the $\bar{p} p \to \phi \phi$ reaction by using the
effective Lagrangian method.

\begin{figure}[htbp]
\begin{center}
\includegraphics[scale=0.58]{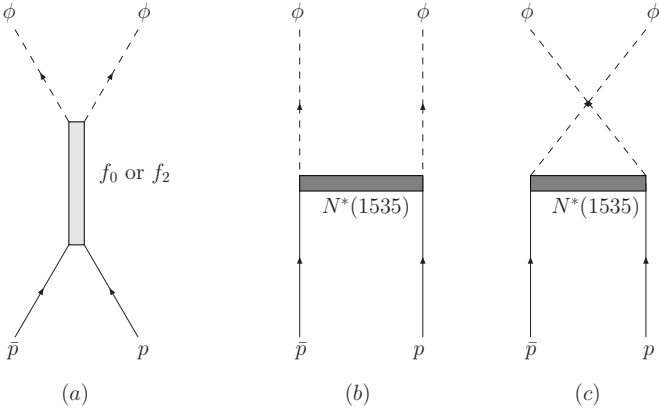}
\caption{Feynman diagrams for $\bar{p} p \to \phi \phi$ reaction.
The contributions from $t$- and $u$-channel $N^*(1535)$ resonance
exchange, and $s$-channel $f_0$ or $f_2$ resonance are considered.}
\label{pbarpdiagram}
\end{center}
\end{figure}

The basic tree level Feynman diagrams for the $\bar{p} p \to \phi
\phi$ reaction are depicted in Fig.~\ref{pbarpdiagram}. In addition
to the ``background" diagrams, such as the $t$-channel
[Fig.~\ref{pbarpdiagram} (b)] and $u$-channel
[Fig.~\ref{pbarpdiagram} (c)] $N^*(1535)$ resonance exchange which
have been considered in the previous calculation~\cite{Shi:2010un},
we  include the $s$-channel diagram [Fig.~\ref{pbarpdiagram} (a)]
through either a scalar meson ($f_0$) or a tensor meson ($f_2$) in
our present calculation.

The invariant scattering amplitudes that enter our model for the
calculation of the total and differential cross sections for the
reaction
\begin{eqnarray}
\bar{p}(p_1,s_1) p(p_2,s_2) \to \phi(p_3,\lambda_1)
\phi(p_4,\lambda_2)
\end{eqnarray}
are defined as
\begin{eqnarray}
-iT_i = \bar{\nu}(p_1,s_1)A_i^{\mu \nu}u(p_2,s_2)
\epsilon^*_{\mu}(p_3,\lambda_1) \epsilon^*_{\nu}(p_4,\lambda_2),
\label{ti}
\end{eqnarray}
where $\nu(p_1,s_1)$ and $u(p_2,s_2)$ are Dirac spinors for
anti-proton and proton, respectively, while
$\epsilon_{\mu}(p_3,\lambda_1)$ and $\epsilon_{\nu}(p_4,\lambda_2)$
are polarization vectors for the $\phi$ mesons. The subscript $i$
stands for the $s$-channel $f_0$ or $f_2$ process, the $t$- and
$u$-channel $N^*(1535)$ resonance exchange.

The explicit expressions for the reduced $A_{N^*(1535)}^{\mu \nu}$
can be found in Ref.~\cite{Shi:2010un}. Here, we only give details
about the $s$-channel $f_0$ and $f_2$  amplitudes, $A^{\mu
\nu}_{f_0}$ and $A^{\mu \nu}_{f_2}$, associated to the diagram of
Fig.~\ref{pbarpdiagram} (a). They are obtained from the following effective
interaction Lagrangian~\cite{Renner:1971mu,Kochelev:1999zf,Oh:2003aw}:
\begin{eqnarray}
{\mathcal L}_{f_0 \bar{p} p}  &=& g_{f_0 \bar{p} p}
\bar{\Psi}_{\bar{p}} f_0 \Psi_{p} \,+{\rm h.c.}, \label{f0nn} \\
{\mathcal L}_{f_0 \phi \phi} &=& g_{f_0 \phi \phi} m_{\phi} \phi_{\mu} \phi^{\mu} f_0 \, ,\label{f0phiphi} \\
{\mathcal L}_{f_2 \bar{p} p} &=& -i \frac{g_{f_2 \bar{p} p}}{m_{N}} \bar{\Psi}_{\bar{p}}
(\gamma_{\mu} \partial_{\nu} + \gamma_{\nu} \partial_{\mu}) \Psi_{p} f^{\mu\nu}_2 \, + {\rm h.c.},  \label{f2nn} \\
{\mathcal L}_{f_2 \phi \phi} &=& g_{f_2 \phi \phi} m_{\phi}
\phi_{\mu} \phi_{\nu} f^{\mu \nu}_2 \, . \label{eq:eqknstar}
\end{eqnarray}

With the above Lagrangians, the reduced
$A_i^{\mu}$ amplitudes in Eq.~(\ref{ti}) can be easily obtained,
\begin{eqnarray}
A_{f_0}^{\mu \nu} &=& -g_{f_0 \bar{p}p} g_{f_0 \phi\phi} m_{\phi} G_{f_0}(q_s) g^{\mu \nu} \, f_s, \label{eq:af0} \\
A_{f_2}^{\mu \nu} &=& i \frac{g_{f_2\bar{p}p} g_{f_2 \phi \phi} m_{\phi}}{m_{N}}
\Big [ \gamma_{\rho} (p_1 - p_2)_{\sigma} + \gamma_{\sigma} (p_1-p_2)_{\rho} \Big ]  \nonumber \\
&&  \times G^{\rho\sigma \mu\nu}_{f_2}(q_s) \, f_s, \,
\label{eq:af2}
\end{eqnarray}
where the propagators for the scalar meson $f_0$ and the tensor mesor $f_2$
are, respectively,
\begin{eqnarray}
G_{f_0}(q_s) &=& \frac{i }{s-M^2_{f_0} + i M_{f_0}\Gamma_{f_0}}, \\
G^{\mu\nu\rho\sigma}_{f_2}(q_s) &=& \frac{i}{s-M^2_{f_2}+i
M_{f_2}\Gamma_{f_2}} P^{\mu\nu \rho\sigma} (q_s),
\end{eqnarray}
and
\begin{eqnarray}
P^{\mu \nu \rho\sigma}(q_s) &=& \frac{1}{2}(\bar{g}^{\mu\rho}\bar{g}^{\nu\sigma} +
\bar{g}^{\mu\sigma}\bar{g}^{\nu\rho}) - \frac{1}{3}\bar{g}^{\mu\nu}\bar{g}^{\rho\sigma} \, , \\
\bar{g}^{\mu\nu} &=& -g^{\mu \nu} + \frac{q^{\mu}_s q^{\nu}_s}{s},
\end{eqnarray}
with $q_s = p_1 + p_2$ the momentum of $f_0$ or $f_2$ and
$s=q^2_s$ the invariant mass square of the $\bar{p} p$ system.

As can be seen from Eqs.~(\ref{eq:af0}) and (\ref{eq:af2}), in the
tree-level approximation,  only the products, $g_{f_0 \bar{p}p}
g_{f_0 \phi\phi}$ and $g_{f_2 \bar{p}p} g_{f_2 \phi\phi}$ enter the
invariant amplitudes. $M_{f_0}$ ($M_{f_2}$) and $\Gamma_{f_0}$
($\Gamma_{f_2}$) are the mass and the total decay width of the $f_0$
($f_2$) meson. We take them as free parameters and determine them by
fitting to the total cross section of the $\bar{p} p \to \phi \phi$
reaction~\cite{Evangelista:1998zg} using MINUIT.

In Eqs.~(\ref{eq:af0}) and (\ref{eq:af2}), we have also included
the relevant off shell form factors~\footnote{We take the following
form factor for $t$- and $u$-channel $N^*(1535)$ ($\equiv N^*$)
resonance exchange as in Ref.~\cite{Shi:2010un}:
\begin{eqnarray}
F_{N^*(1535)} &=& \frac{\Lambda^2_{N^*} -M^2_{N^*}}{\Lambda^2_{N^*}
- q^2_{N^*}}, \nonumber
\end{eqnarray}
with $q^2_{N^*}$ the 4-momentum of the exchanged $N^*(1535)$
resonance. In general, the cutoff parameter $\Lambda_{N^*}$ for
$N^*(1535)$ resonance should be at least a few hundred MeV larger
than the $N^*(1535)$ mass, and thus in the range of $2$ to $4$
GeV.} for $f_0$ and $f_2$ mesons. We adopt here the common scheme
used in many previous works,
\begin{eqnarray}
&& f_s =\frac{\Lambda^4_i}{\Lambda^4_i+(s - M_i^2)^2}, \quad i= f_0,
f_2 \, .  \label{sff}
\end{eqnarray}
The cutoff parameters, $\Lambda_{f_0}$ and $\Lambda_{f_2}$, are
constrained between $0.6$ and $1.2$ GeV. This way, we can reduce the
number of free parameters.

\section{Numerical results and discussion}

The differential cross section for $\bar{p} p \to \phi \phi$
reaction at the center of mass ($\rm c.m.$) frame can be expressed as
\begin{equation}
{d\sigma \over d{\rm cos}\theta}={1\over 64\pi s}{
|\vec{p}_3^{\text{~c.m.}}| \over |\vec{p}_1^{\text{~c.m.}}|} \left (
{1\over 4}\sum_{s_1,s_2,\lambda_1,\lambda_2}|T|^2 \right ),
\label{eq:pipdcs}
\end{equation}
where $\theta$ denotes the angle of the outgoing $\phi$ meson
relative to the beam direction in the $\rm c.m.$ frame, while
$\vec{p}_1^{\rm{~c.m.}}$ and $\vec{p}_3^{\text{~c.m.}}$ are the
3-momentum of the initial $\bar{p}$ and final $\phi$ meson.

First, by including the contributions from the $s$-channel scalar
meson $f_0$~\footnote{In general, we should study the role of the
scalar meson and tensor meson together. However, because of the
limitation of the experimental measurements and scarcity of the
information about the relevant mesons, hence, we separately study
them in this work.} and $t$- and $u$-channel $N^*(1535)$ resonance
(corresponding to $T= T_{f_0} + T_{N^*(1535)}$), with fixed cutoff
parameters $\Lambda_{f_0}$ and $\Lambda_{N^*(1535)}$, we perform a
$\chi^2$ fit (Fit I) to the total cross section data for $\bar{p} p
\to \phi \phi$~\cite{Evangelista:1998zg}. There are a total of $20$
data points.

By constraining the value of the cutoff parameter $\Lambda_{f_0}$
between $0.6$ and $1.2$ GeV and $\Lambda_{N^*(1535)}$ around $3.0$ GeV
based on the results of Ref.~\cite{Shi:2010un}, we obtain a minimal
$\chi^2/\mathrm{d.o.f.} = 2.1$ with $\Lambda_{f_0} = 0.6$ GeV and
$\Lambda_{N^*(1535)} = 3.05$ GeV. The fitted parameters are:
$g_{f_0 \bar{p}p} g_{f_0 \phi \phi} = 0.45 \pm 0.08$, $M_{f_0} =
2174 \pm 3$ MeV, and $\Gamma_{f_0} = 167 \pm 27$ MeV.

Second, instead of a scalar meson, we study the case of a tensor
meson $f_2$ in the $s$-channel and $t$- and $u$-channel $N^*(1535)$
resonance (corresponding to $T= T_{f_2} + T_{N^*(1535)}$), and we
perform a second $\chi^2$ fit (Fit II). In this case, we get a
minimal $\chi^2/\mathrm{d.o.f.} = 1.4$ with $\Lambda_{f_2} = 0.65$
GeV and $\Lambda_{N^*(1535)} = 3.05$ GeV. The fitted parameters are:
$g_{f_2 \bar{p}p} g_{f_2 \phi \phi} = -0.12 \pm 0.02$, $M_{f_2} =
2192 \pm 4$ MeV, and $\Gamma_{f_2} = 177 \pm 30$ MeV.

Based on the value of the $\chi^2/\mathrm{d.o.f.}$, Fit II is
preferred to Fit I. It seems to indicate that the $\bar{p} p \to
\phi \phi$ reaction is dominated by the exchange of a strange tensor
meson with quantum number $J^{PC} = 2^{++}$ in the $s$-channel, in
agreement with the study of Ref.~\cite{Evangelista:1998zg}. In
addition, a partial-wave analysis of the $\pi^- p \to \phi \phi n$
reaction shows that the $\phi \phi$ system is dominant by two
$J^{PC} = 2^{++}$ states~\cite{Etkin:1982bw}, one an $S$ wave and
the other a $D$ wave. The mass of the $S$ wave state is $M = 2160
\pm 50$ MeV, with a decay width $\Gamma = 310 \pm 70$ MeV. The mass
is in agreement with our fitted result for the tensor meson.

\begin{figure}[htbp]
\begin{center}
\includegraphics[scale=0.45]{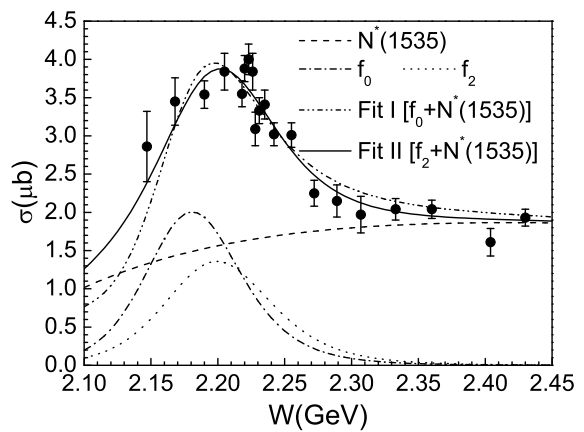}
\caption{Total cross sections for $\bar{p} p \to \phi \phi$
reaction. The experimental data are taken from
Ref.~\cite{Evangelista:1998zg}. The curves are the contributions
from $s$-channel $f_0$ (dash-dotted) and $f_2$ (dotted), $t$- and
$u$-channel $N^*(1535)$ resonance (dashed), and the total results of
Fit I (dash-dot-dotted) and Fit II (solid).} \label{pbarptcs}
\end{center}
\end{figure}

Next, we show the corresponding fitted results for the total cross
sections in Fig.~\ref{pbarptcs}, in comparison with the experimental
data from Ref.~\cite{Evangelista:1998zg}. In Fig.~\ref{pbarptcs},
the dashed curve stands for the contributions from the $t$- and
$u$-channel $N^*(1535)$ resonance, and the dash-dotted and dotted
lines stand for the contributions from the $s$-channel scalar meson
$f_0$ and tensor meson $f_2$, respectively, while the total results
of Fit I and Fit II are shown by dash-dot-dotted and solid curves.
From Fig.~\ref{pbarptcs}, one can see that the experimental total
cross section can be described fairly well by including the
contributions from both the $N^*(1535)$ resonance and  the scalar
meson $f_0$ or tensor meson $f_2$. The contributions from
$N^*(1535)$ resonance dominates above $W = 2.25$ GeV, while the bump
structure around $W = 2.2$ GeV can be well reproduced by considering
the contributions from the strange mesons $f_0$ and $f_2$.

\begin{figure}[htbp]
\begin{center}
\includegraphics[scale=0.42]{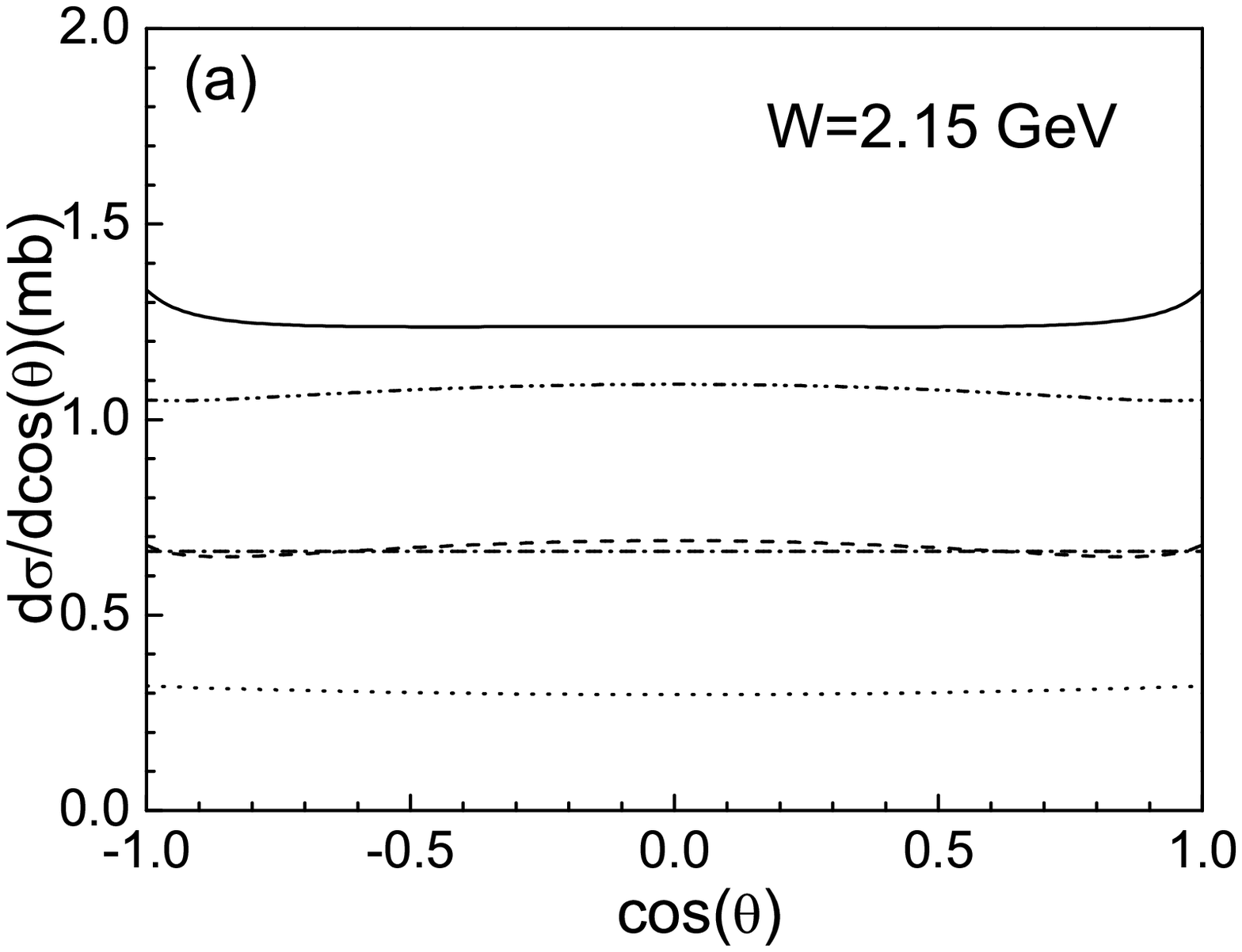}
\includegraphics[scale=0.42]{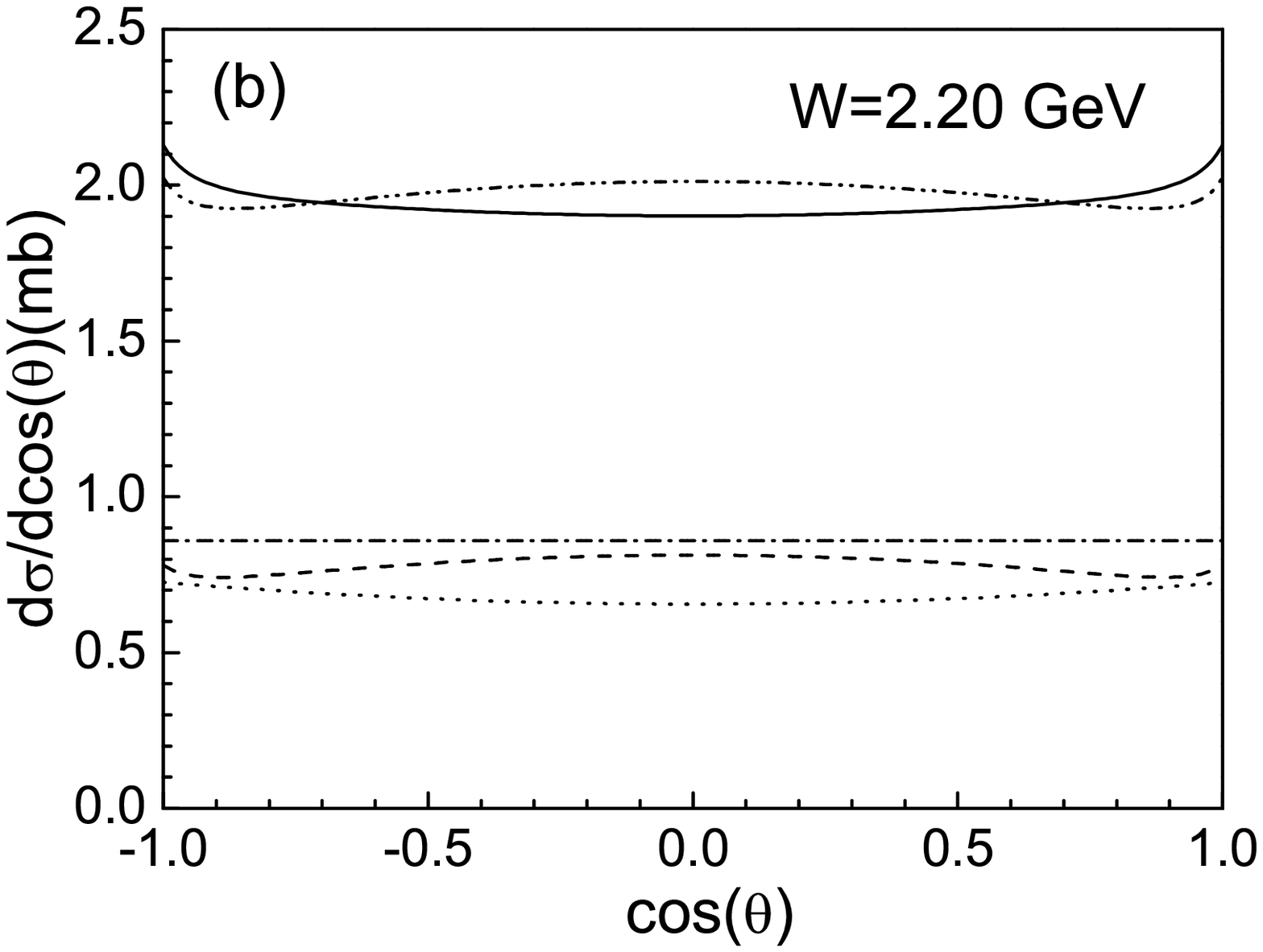}
\includegraphics[scale=0.42]{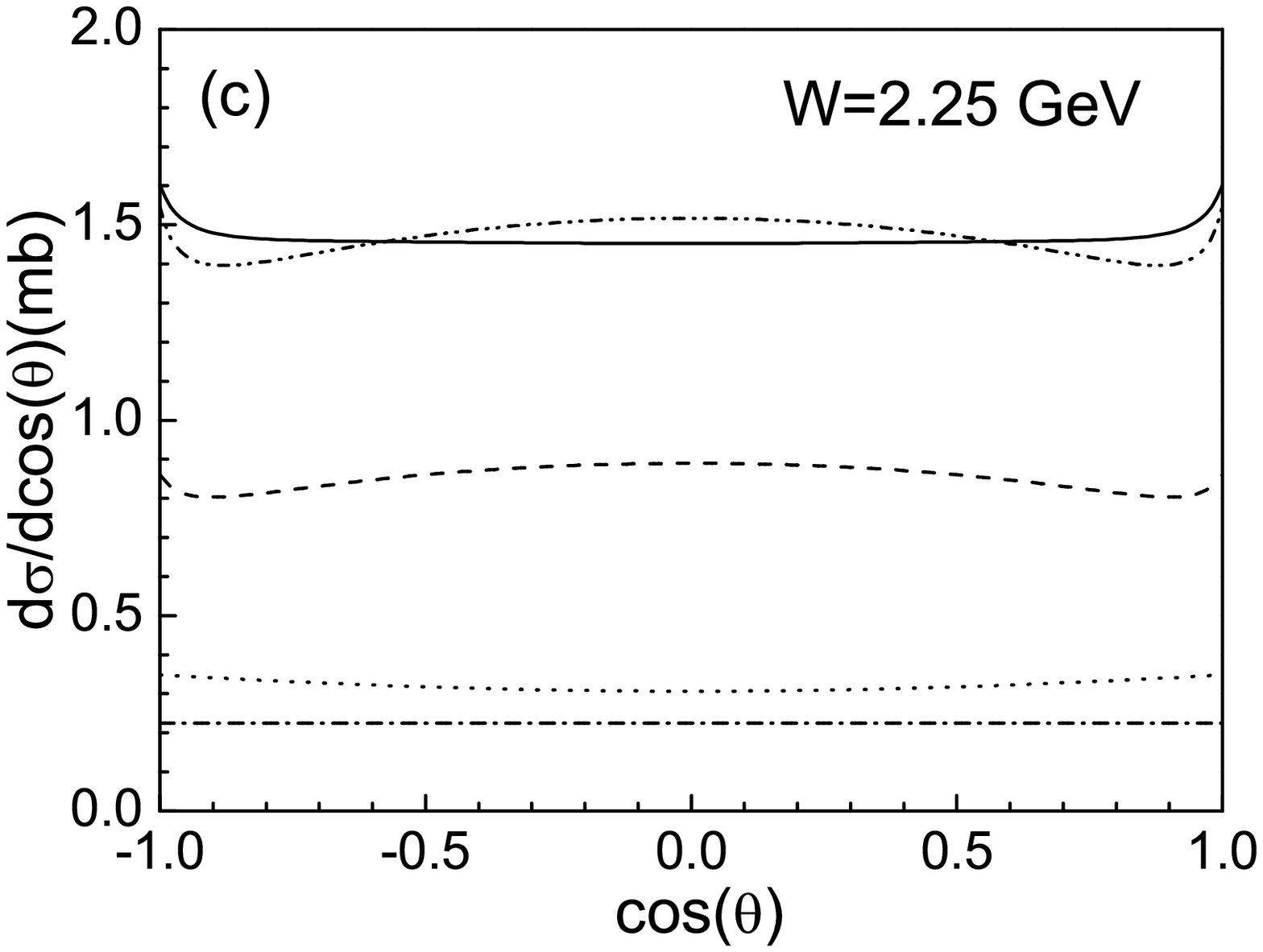}
\caption{Differential cross sections for $\bar{p} p \to \phi \phi$
reaction. The curves are the contributions from $s$-channel scalar
meson $f_0$ (dash-dotted) and tensor meson $f_2$ (dotted), $t$- and
$u$-channel $N^*(1535)$ resonance (dashed), and the total results of
Fit I (dash-dot-dotted) and Fit II (solid).} \label{pbarpdcs}
\end{center}
\end{figure}

With the above fitted parameters, the corresponding differential
cross sections for $\bar{p} p \to \phi \phi$ reaction at the energy
around the fitted masses of $f_0$ and $f_2$, $W = 2.15$ GeV, $W =
2.20$ GeV, and $W = 2.25$ GeV, are shown in Fig.~\ref{pbarpdcs}(a),
Fig.~\ref{pbarpdcs}(b), and Fig.~\ref{pbarpdcs}(c), respectively.
From Fig.~\ref{pbarpdcs}, we see that the shapes of the angular
distributions are similar, mainly because both the scalar meson and
the tensor meson decay to $\phi \phi$ in the $s$-wave. But, there
are still a little bit difference in the two cases, especially for
the energies of $W = 2.20$ GeV and $W = 2.25$ GeV, because the
production of a scalar meson $f_0$ from $\bar{p} p$ is in $s$-wave,
while the $\bar{p}p$ to the tensor meson $f_2$ is in the $D$-wave.
These predictions can be checked by future experiments.

\section{Summary}

In this paper, we have phenomenologically reanalyzed the $\bar{p} p
\to \phi \phi$ reaction within an effective Lagrangian approach and
the isobar model. In addition to the ``background" contributions
from $t$- and $u$-channel $N^*(1535)$ resonance, we studied the role
of scalar meson ($f_0$) and tensor meson ($f_2$) in the $s$-channel.
Unfortunately, the information about the $f_0$ and $f_2$ meson with
mass around $2.2$ GeV is scarce~\cite{pdg2012}. Thus, in the present
work, we have taken the masses, the total decay widths, and the
coupling constants, $g_{f_0\bar{p}p}g_{f_0 \phi \phi}$ and
$g_{f_2\bar{p}p}g_{f_2\phi\phi}$ as free parameters, and we fitted
them to the experimental data on the $\bar{p} p \to \phi \phi$
reaction in Ref.~\cite{Evangelista:1998zg}. The fitted results are:
$M_{f_0} = 2174 \pm 3$ MeV, $\Gamma_{f_0} = 167 \pm 27$ MeV,
$M_{f_2} = 2192 \pm 4$ MeV, and $\Gamma_{f_2} = 177 \pm 30$ MeV. The
fitted results are shown that the $\bar{p} p \to \phi \phi$ reaction
is dominated by the exchange of a strange tensor meson with quantum
number $J^{PC} = 2^{++}$ in the $s$-channel, which is in agreement
with the previous analysis~\cite{Evangelista:1998zg,Etkin:1982bw}.
In this respect, we have shown how the experimental measurements for
the $\bar{p} p \to \phi \phi$ reaction could lead to valuable
information on scalar and tensor mesons with masses around $2.2$
GeV.

Finally, we would like to stress that due to the important role
played by the resonant contribution in the $\bar{p} p \to \phi \phi$
reaction, the bump structure around $W = 2.2$ GeV in the total cross
section can be well reproduced, and more accurate data on this
reaction can be used to improve our knowledge on the strange mesons
$f_0$ and $f_2$, which is at present poorly known. This work
constitutes a first step in this direction.

\section*{Acknowledgments}

We would like to thank Jun Shi and Xu Cao for useful discussions.
This work is partly supported by the Ministry of Science and
Technology of China (2014CB845406), the National Natural Science
Foundation of China under grants: 11105126, 11375024 and 11175220.
We acknowledge the one Hundred Person Project of Chinese Academy of
Science (Y101020BR0).

\end{document}